\documentstyle[12pt,aaspp4]{article}

       \def\teff{$T_{\mbox{{\small eff}}}$}
        \def\logg{$\log(g)$}
       
\begin{document}       
     
\title{Phase Dependent Spectroscopy of Mira Variable Stars} 
  
\author{Michael W. Castelaz\altaffilmark{1,}\altaffilmark{2}, 
Donald G. Luttermoser} 
\affil{Department of Physics, Box 70652, East Tennessee State University,  
Johnson City, TN 37614 and Southeastern Association for Research in Astronomy} 

\author{Daniel B. Caton}
\affil{Department of Physics and Astronomy, Dark Sky Observatory, Appalachian
 State University, Boone, NC 28608} 
 
\author{Robert A. Piontek\altaffilmark{1,}\altaffilmark{3}} 
\affil{University of Maryland, Department of Astronomy, 
College Park, MD  20742}
   
\altaffiltext{1}{Visiting Astronomer, Dark Sky Observatory, Appalachian State 
University} 

\altaffiltext{2}{Also at The Pisgah Astronomical Research Institute}  
   
\altaffiltext{3}{Participated in the Summer 1998 Southeastern Association
 for Research in Astronomy Research Experience for Undergraduates Program 
Sponsored by the National Science Foundation} 
 
\authoremail{mwc@etsu.edu \& lutter@etsu.edu \& 
catonb@appstate.edu \& rpiontek@ra.astro.lsa.umich.edu}   
 
\begin{abstract}

Spectroscopic measurements of Mira variable stars, as a function of phase,   
probe the stellar atmospheres and underlying pulsation mechanisms.  For 
example, measuring variations in TiO, VO, and ZrO with phase can be used to 
help determine whether these molecular species are produced in an extended 
region above the layers where Balmer line emission occurs or below this 
{\it shocked} region.  Using the same methods, 
the Balmer-line {\it increment}, where the strongest
Balmer line at phase zero is H$\delta$ and not H$\alpha$ can be measured and 
explanations tested, along with another peculiarity, the absence 
of the H$\epsilon$ 
line in the spectra of Miras when the other Balmer lines are strong.  We 
present new spectra covering the spectral range from 6200~\AA\ to 9000~\AA\ of 
20 Mira variables.  A relationship between variations in the \ion{Ca}{2} IR 
triplet and H$\alpha$ as a function of phase support the hypothesis that 
H$\epsilon$'s observational characteristics result from an interaction of
H$\epsilon$ photons with the \ion{Ca}{2} H line.  New periods and  epochs
of variability are also presented for each star.
 
\end{abstract}         
   
         
\section{Introduction}         
         
Mira-type variable stars are large, cool stars whose visual light variations 
exceed 2.5 magnitudes over periods from 150 days to $\sim$500 days.  The light 
curves of Mira variables depend on the surface temperature, radius, and 
opacity, all which vary as the star pulsates.  These pulsations extend the 
atmosphere beyond that of the hydrostatic equilibrium configuration and 
enhance mass loss in these stars (Maciel 1977; Willson \& Hill 1979; 
Bertschinger \& Chevalier 1985; Bowen 1988; Fleischer et~al.\ 1992, 1995).  As 
a result, Mira variables are an important component in seeding the 
interstellar medium with C, N, and O. 
 
These stars are located on the asymptotic giant branch, 
a transitional phase in 
stellar evolution.  Photometric and spectroscopic measurements of their light 
curves provide a means to probe the stellar atmospheres and underlying 
pulsation mechanisms occurring during this stellar phase.  In the near
infrared, the spectra of Mira stars are dominated by the TiO $\gamma$ system, 
the VO $\gamma$ system and ZrO (Wing 1967).  The TiO features are thought to be
produced in a layer somewhat far from the photosphere (Gillet 1988).  Haniff 
et~al.\ (1992) present optical aperture synthetic images of the photosphere
of {\it o} Ceti 
at 6500~\AA, 7007~\AA, and within a TiO bandhead at 7099~\AA, with the star 
phase $\sim$0.94.  They find asymmetry in the images, with the TiO image one 
and a half times larger than the photospheric images.  Also, narrowband speckle
interferometric measurements taken in the TiO 7120~\AA\ bandhead and outside at
7400~\AA\ by Labeyrie et al.\ (1977) shows that the diameters of R~Leo and 
$o$~Cet are twice as large in the TiO feature than outside of it.  This 
demonstrates that a model atmosphere, based on the spectra observed over a TiO 
bandpass, provides parameters such as  \teff\ and \logg\ 
in an atmospheric layer far from the photosphere.
 
 Joy (1926) presents a comprehensive phase dependent 
spectroscopic (35 \AA/mm) study of a prototype Mira 
variable, {\it o} Ceti.  Analyzing 131
spectra  taken over a ten year period, Joy describes several 
important characteristics.  Briefly, the spectra of {\it o} Ceti show 
that TiO bands vary with magnitude, hydrogen emission lines appear 
with greatest intensity at or shortly after maximum visual brightness
(phase zero), 
ionized iron emission lines are observed at maximum, and the temperature is
estimated to vary from 1800~K to 2300~K.  Absorption lines 
(including iron, vanadium, chromium, manganese, calcium, and magnesium)
were used to measure a variation in radial velocity.  The maximum 
positive velocity occurs at phase zero, and greatest blueshift at minimum
light. Later, Joy (1954) took 88 
spectrograms of {\it o} Ceti at a higher spectral resolution 
(typically 10.3 \AA/mm) and confirmed that maximum velocity 
of recession occurs soon after visual maximum.  He attributed these
results to a pulsational mechanism, and suggested the possibility of
shocks.   

Perhaps one of the most interesting characteristics of Mira spectra is the
strong hydrogen Balmer line emission that is seen throughout much of the 
pulsation cycle.  As Pickering (1887) first noticed in spectra of {\it o} Ceti,
and later described in detail by Joy (1926, 1947, 1954), the 
hydrogen Balmer emission line flux in relation to the
nearby {\it photospheric} (i.e., pseudocontinuum) flux is unique in the
oxygen-rich (M-type) Mira spectra: Balmer H$\alpha$ emission is typically 
weaker than H$\beta$ which in turn is weaker than H$\gamma$ near peak visual 
brightness.   H$\delta$ is seen as the strongest Balmer emission line at 
phase zero.  Lines higher in the series 
(i.e., towards 
shorter wavelengths), are weaker.  This {\it Balmer increment}
(i.e., $F_{H\alpha} < F_{H\beta} < F_{H\gamma} < F_{H\delta}$) is just opposite
of what would be expected, H$\alpha$ having the largest oscillator strength, 
should be stronger than H$\beta$ and the higher order Balmer lines should be 
weaker down the line (i.e., one should see a Balmer line {\it 
decrement}), assuming these lines all form in the same region of the atmosphere
(i.e., similar $T$ and $P$).  Meanwhile, in S-type and carbon-star (N-type)
Miras, the strength of the Balmer lines approximately follow their
expected  respective
oscillator strengths (Merrill 1940).

For years, this Balmer-line increment in the M-type Miras has been attributed 
to TiO absorption which hide H$\alpha$, H$\beta$, and H$\gamma$ fluxes (Merrill
1940; Joy 1947; Gillet 1988), although there is some 
debate of the extent that this or 
other molecular absorption has on the H$\alpha$ line 
(Gillet, Maurice, \& Baade 
1983; Gillet et~al.\ 1985).  Recently another explanation has been given for
this Balmer increment: NLTE radiative transfer calculations of hydrodynamic 
models representative of Mira variables (Bowen 1988) suggest that the Balmer 
increment results from radiative transfer effects in the hydrogen lines 
themselves when formed in a shocked atmosphere (Luttermoser \& Bowen 1992; 
Luttermoser, Bowen, \& Willson 2000, in preparation).  
In these calculations, H$\alpha$, having
the highest optical depth, forms just in front of the innermost shock.
H$\beta$ then forms a little deeper, due to its lower optical depth, and
H$\gamma$ deeper still.  The optical depth of H$\delta$ causes it to arise from
the hottest part of the shock, hence extends higher above the continuum then
the longward Balmer lines.  Then as one goes to higher-order lines in the
series, the opacity in these lines is not high enough in the shock for these
lines to form there --- we see through the shock at these transitions.  As
such, this increment may be giving us information of the shock thickness for 
Miras and may indicate that the shock structure of the S-type and N-type Miras 
are fundamentally different than the M-type Miras.  Future NLTE radiative 
hydrodynamic models of Miras with different C/O ratios are needed to see if 
this is the case.  It is likely that a combination of both processes (i.e., TiO
absorption and NLTE radiative transfer effects) are responsible for this Balmer
line increment.
                 
Another striking feature is the weakness (and often absence) 
of the H$\epsilon$ 
line (3970.074~\AA) in the Balmer series near maximum visual brightness  (see 
Gillet 1988).  Merrill (1940) noted this and suggested this weakness a result
of the interaction between the H$\epsilon$ transition and the Ca~II H line 
(3968.470~\AA) wing.  Castelaz \& Luttermoser (1997, hereafter CL97) concur 
with this suggestion.  Briefly, the H$\epsilon$ photons may be scattered by the
Ca~II H line out to IR wavelengths via the 8662~\AA\ line.  Of the three lines
in the Ca~II IR triplet, the 8498~\AA\ and 8542~\AA\ lines share the same upper
level of the Ca~II K line, whereas the 8662~\AA\ line shares the same upper
level as the Ca~II H line.  As such, if H$\epsilon$ photons are being scattered
by Ca~II H, the 8662~\AA\ will show variations independent of the other two
Ca~II IR triplet lines and can be tested by monitoring the strength of the 
absorption of this 8662~\AA\ line as compared to the 8498~\AA\ and 8542~\AA\
lines as a function of phase in the Mira stars.

Ca~II IR triplet observations of Miras have been carried out by Contadakis \& 
Solf (1981) and Gillet et~al.\ (1985).  Unfortunately, Contadakis \& Solf 
(1981) only made one observation of the $\lambda$8662 line in their monitoring
program of S-type Miras.  They observed the $\lambda$8498 and $\lambda$8542 in 
emission in many of the stars in their sample near phase 0 and as absorption
lines at other phases.  So, we cannot use their data to test our proposed
hypothesis.  Gillet et~al.\ (1985) observed P~Cygni profiles in the Ca~II IR
triplet lines near phase 0 in the hot Mira variable, S~Car.  
Once again, only one
spectrum of the 8662~\AA\ line was obtained in their sample of Ca~II spectra.
CL97 present a set of phase dependent spectra taken especially to address this
problem of H$\epsilon$ photon scattering by Ca~II.  Spectra of seven Mira
variables were taken at different phases and suggest a possible anticorrelation
between H$\alpha$ emission and Ca~II $\lambda$8662 absorption.  Assuming that
the H$\epsilon$ line strength variations are in phase with H$\alpha$, then the 
apparent anticorrelation between the strength of the H$\alpha$ emission line 
and the strength of the Ca~II $\lambda$8662 line suggests that a {\it 
fluorescence} is taking place in the Ca~II $\lambda$8662 line with H$\epsilon$ 
serving as the pump through the Ca~II H line.  In this paper we present phase 
dependent spectra of twenty Mira variables to further explore the 
anticorrelation of Ca~II $\lambda$8662 absorption with H$\alpha$ emission. 
 
\section{Observations}       

 The twenty stars for which spectra are presented, 
their equatorial coordinates, 
mean spectral types, and visual maxima and minima, taken from the SIMBAD 
database are presented in Table 1.  Also in Table 1 are new ephemerides of 
these Mira stars.  We calculated the ephemrides from AAVSO 
light curves measured within the previous eight years.  
The curves were fit with a linear combination of
sine and cosine functions from which a new period and JDO were determined 
for each star.  
         
Spectra of Mira variable stars taken between February 1997 and November 1998 
using a low resolution spectrograph.  Since we are interested in only
monitoring absorption and emission line strengths for this program,
low-resolution spectra are all that is required.  The spectrograph was used at
both the Southeastern Association for Research in Astronomy (SARA) 0.9-m
telescope at Kitt Peak, and Appalachian State University's Dark Sky Observatory
(DSO) 0.45-m telescope located near Boone, NC.  A converter lens was used at
both sites to convert the respective telescope f-ratio to about f/11 for the
spectrograph.

The spectrograph was configured with a 600 g/mm grating.  The slit width was 
100 $\mu$m.  At the SARA 0.9-m telescope, the slit width is 3 arcseconds, and 
at the DSO 0.45-m telescope, the slit width is 4 arcseconds.  At both sites, 
the slit was parallel to the hour angle.  A cooled 768 x 512 CCD camera with 9 
$\mu$m square pixels was used to record the spectra.  The linear dispersion is 
1.08~\AA\ per pixel, covering 768 pixels, or 829~\AA\ on the CCD.  The spectral
resolution was measured to be 2.4~\AA.  By rotating the grating up to four 
times per star, spectra were taken from about 6200~\AA\ to 9000~\AA\ and 
include H$\alpha$, TiO, VO, and the Ca~II infrared triplet lines.  Integration 
times were adjusted to achieve a signal-to-noise greater than 100 for most 
spectra, except for the May 1997 spectra of W Her, T Her, and W Lyr, and the 
May 1998 spectrum of T Her.
         
Table 2 gives the log of observations for  each star, which includes dates of 
observation, phase of the variable,  approximate visual magnitude, integration
time, and observing site.  The phases listed in Table 2 were determined from 
the ephemerides given in Table 1, and refer to the visual phases, with phase 
zero corresponding to maximum visual  brightness.  The visual magnitudes have 
been obtained from curve fits to the light curves from the AAVSO database.
 
Dark frames and flat frames were taken for flat fielding purposes.  Spectra of 
neon lamp emission were taken simultaneously with the stellar spectra and used
for wavelength calibration.  We flat-fielded the images and extracted the 
spectra using MIRA software.  The extracted spectra were wavelength calibrated 
using the spectrum of neon superimposed on the CCD frame with the stellar 
spectrum (Crowe, Heaton, \& Castelaz 1996).  

\section{Results}         
         
The spectra for the Mira variable stars are shown in Figure 1.  
The wavelengths 
of TiO, VO, ZrO, the Ca~II IR triplet,  H$\alpha$, and terrestrial oxygen are 
marked above each set of spectra.  Their wavelengths were identified in CL97. 
The ZrO and VO absorption overlap at 6574~\AA\ and 6578~\AA, and VO and TiO 
absorb at 7865~\AA\ and 7861~\AA, respectively.  Due to the low dispersion of 
our spectra, these features are blended.  The appearance of a relatively narrow
feature near 8230~\AA\ is seen in some of the spectra - an $o$~Cet 
spectrum taken 1 November 1997 (phase 0.80), and spectra 
taken 26-28 May 1997 of 
R~Leo (0.49), V~CVn (0.65), R~Boo (0.11), R~Ser (0.26), RS~Her (0.75), and 
W~Lyr (0.94).  This feature is due to terrestrial H$_{2}$O at 8227~\AA\ 
(Turnshek et al.\ 1985) and is an effective measure of the relative humidity in
the air.  

\subsection{Radiative Transfer in the Ca~II Ion}

The transfer of radiation in Ca~II ion is very complicated.  Beside 
H$\epsilon$ photons affecting the level populations of the 3d~$^{2}$P$_{1/2}$
state through the Ca~II H line, the hydrogen Lyman-$\alpha$ line lies just
shortward of the Ca~II ionization edge of the 3d~$^{2}$D$_{3/2}$ (at 
1218.1~\AA) and the 3d~$^{2}$D$_{5/2}$ (at 1219.0~\AA) states.  These two
states are the lower levels of the Ca~II IR triplet lines and are metastable.
If  Lyman-$\alpha$ is a strong emission feature, 
and the Ca~II $^{2}$D continuous 
opacity forms in a region of the atmosphere where the Lyman-$\alpha$ line is
not yet in detailed balance, then 
Lyman-$\alpha$ photons may influence the Ca~II IR
triplet lines as well.   Note that no observations 
have yet been made in the far-UV for Miras.

Line center of Lyman-$\alpha$ lies 2.4~\AA\ shortward of the $^{2}$D$_{3/2}$
ionization edge and 3.3~\AA\ shortward of the $^{2}$D$_{5/2}$ edge.  So, 
much of the Lyman-$\alpha$ emission profile can ionize Ca~II out of the
metastable state.  The lower level of the Ca~II 8662~\AA\ line transition is
the $^{2}$D$_{3/2}$ state whose ionization edge lies slightly closer to
Lyman-$\alpha$ than the $^{2}$D$_{5/2}$ edge.  The following question arises:
Will Lyman-$\alpha$ photons affect the level densities of the two 3d~$^{2}$D
states differently?  To answer this question, we ran a few atmospheric models
with an arbitrarily located 10,000~K shock through the LTE stellar atmosphere
code ATLAS (Kurucz 1970; Brown et~al.\ 1989).  Although it is likely that NLTE
effects will dominate the level and ion densities in the atomic and molecular
species (Luttermoser \& Bowen 1992; Luttermoser, Bowen, \& Willson 2000,
in preparation), these
LTE runs are performed to merely determine the variation of the Ca~II $^{2}$D
continuous opacity across the Lyman-$\alpha$ profile.  It also should be 
pointed out that in regions of the atmosphere where LTE
no longer applies, the assumption of radiative and hydrostatic equilibria are
no longer valid either in these pulsating giant stars.  In fact, a Mira star
has numerous shocks propagating through its atmosphere at any given time as has
been shown by Bowen (1988) and more recently by H\"{o}fner, et~al.\ (1998) and
Loidl, et~al.\ (1999).  Willson (2000) gives a very detailed review of all the
dynamic modeling that has been performed on these pulsating stars and discusses
the problems of carrying out NLTE radiative transfer in such a dynamic
atmosphere.

We sampled atmospheric
depths in front of the shock, in the shock, behind the shock, and deep in the
photosphere where the continuum reaches optical depth unity in this region of
the spectrum.  We found that the continuous opacity from the Ca~II 3d~$^{2}$D
ionization remained constant (from both $J$ sub-levels) to within 0.15\% from
the location of the edges through 1210~\AA, which should include most of the
Lyman-$\alpha$ emission profile.  The fact that the continuous opacity from
Ca~II remains constant across the Lyman-$\alpha$ profile indicates that
Lyman-$\alpha$ will not preferentially affect the number density in the
$^{2}$D$_{3/2}$ level as compared to the $^{2}$D$_{5/2}$ level ---
photoionizations of Ca~II due to Lyman-$\alpha$ photons will affect the
strengths of three Ca~II IR lines in a similar fashion.  Therefore, any
variation in the 8662~\AA\ line that is not seen in the other lines must result
from some process other than Lyman-$\alpha$ photoionizations.

\subsection{The H$\alpha$ Emission Line and Ca~II IR Triplet} 

We are interested in the strength of the Ca~II IR triplet  compared to the
strength of H$\alpha$ as a function of phase, since we are assuming that
variations in H$\epsilon$ will mimic variations in H$\alpha$.  
The Ca~II IR triplet lines are not strong, as expected for stars later
than M0 (Zhou 1991).  A total of twenty-seven spectra of fifteen stars in our
sample span the wavelength range from H$\alpha$ through the Ca~II IR triplet.
The remaining spectra are missing either the H$\alpha$ or the 
Ca~II IR triplet regions of the spectrum because the spectrograph 
grating was not rotated sufficiently during
observation to cover those parts of the the spectrum.   
 
The observations which show obvious H$\alpha$ emission features include U~Ori 
(phase 0.29), R~Leo (0.34), V~CVn (0.12), R~CVn (0.09 and 0.17), 
R~Ser (0.26), W~Lyr 
(0.94), and $\chi$~Cyg (0.04).  At the 
same phases, the Ca~II $\lambda$8662 is  seen in emission in R~Leo, R~Ser, 
W~Lyr, and $\chi$~Cyg, whereas the $\lambda$8498 and $\lambda$8542 lines stay
in absorption or are not apparent.  Ca~II $\lambda$8662 is not seen in emission
in any other spectra that we took, only in those that show H$\alpha$ emission.
The Ca~II IR triplet can be seen in absorption in the remaining twenty--three 
spectra (although in some cases weakly), except for U~Ori (phase 0.01), 
S~CrB (0.48),
W~Her (0.86), and T~Her (0.15) which do not appear to have any type 
of Ca~II IR triplet features.

At this point, we call attention to three of the hydrogen 
Paschen lines which lie close to each Ca~II IR triplet 
line: 8502.4~\AA\ (4.4~\AA\
redward from the Ca~II $\lambda$8498.0 line), 8545.3~\AA\ (2.8~\AA\ redward of
the $\lambda$8542.1 line), and 8665.0~\AA\ (2.8~\AA\ from the $\lambda$8662.1
line), the Pa13, Pa12, and Pa10 lines respectively.  It has been shown by
Gillet et~al.\ (1985) that even though the Pa-$\delta$ line is in emission in
the spectrum of the Mira variable S~Car, the higher order Paschen lines near
the Ca~II IR triplet are neither seen in absorption nor emission, analogous to
the Balmer lines.  As such, it is unlikely that these higher order Paschen
lines are affecting the Ca~II lines in our spectra.

\subsection{The TiO Bands and H$\alpha$ Emission} 
 
In addition to the Ca~II IR triplet, we are interested in comparing the TiO
$\gamma$ system molecular features to H$\alpha$ emission as a function 
of phase.  A qualitative comparison of H$\alpha$ emission with the molecular
features can be made for U~Ori, R~Leo, R~CVn, R~Ser, W~Lyr, and $\chi$~Cyg, 
stars with phase dependent spectra that also show the H$\alpha$ emission line 
in at least one spectrum.  Weak H$\alpha$ emission is seen in U~Ori at phase 
0.29 on 15 Feb 1998.  However on the previous pulsation cycle H$\alpha$
emission is not obvious at either phases 0.30 or 0.01, 
although there may some very weak emission.  Stronger H$\alpha$ emission is
seen in R~Leo (phase 0.34), R~CVn (0.09, 0.17, and 0.39), R~Ser (0.26), 
W~Lyr (0.94), and $\chi$~Cyg
(0.04).  As mentioned earlier in the paper, H$\alpha$ is notorious
for being observed as a weak emission feature when H$\beta$, H$\gamma$, and
H$\delta$ are strong.  Surprisingly though, H$\alpha$ emission was not seen in
various spectra where we would expect to find it: R~Tau (phase 0.16 and 0.99), 
R~Leo (0.16 and 0.25), V~CVn (0.12,
0.27, and 0.06), R~CVn (0.20), V~Boo (0.12), and T~Her (0.15).

Merrill (1940) and Joy (1926) report that the TiO bands are 
regularly stronger at minimum 
than at maximum light in Mira variables.  To measure this trend, we 
checked the variability of the TiO feature at 7054~\AA\ (bandhead) with respect
to a portion of the flux uncompromised by TiO, VO, and ZrO.   We integrated the
flux of each spectrum in the 6995-7045~\AA\ wavelength band (non-TiO) and the
7060-7110~\AA\ band (TiO), each 50~\AA\ wide.  We then divided the integrated
flux of the TiO band by the non-TiO band.  By doing this, any scattered light
that may be in the spectra are effectively canceled out.  Since our data set 
contains warm oxygen-rich (earlier than M6), cool oxygen-rich (M6-M8), and one
MS Mira sampled sporadically over various phases, we only include the cool
oxygen-rich Miras in Figure 2, which graphs the above mentioned flux ratio as a
function of phase.  The M6-M8 stars are selected here in order to minimize the
spread of effective temperatures at maximum light which will influence the
strength of the TiO bands.  Ideally, one would want many observations of each
star over a single pulsation cycle.  However, with the limited sample we have,
we feel that we can get an approximate test of TiO variation with respect to
phase.  There appears to be no obvious trend in
variations in the TiO 7054~\AA\ band flux with respect to the non-TiO band
flux.  Due to this observation, variations seen in H$\alpha$ as a function of
phase must primarily result from variations in the H$\alpha$ emission itself
and not to variations in overlying TiO absorption.  As a result, using
H$\alpha$ flux variations as a proxy to variations in the intrinsic H$\epsilon$
flux is valid from this analysis.

\section{Discussion}         
         
We wish to test the idea that the apparent lack of H$\epsilon$ emission at
3970~\AA\ when the other Balmer lines are strong emission features is
anticorrelated with the strength of the Ca~II absorption line at 8662~\AA.  As
reported in the Introduction, this anticorrelation results from H$\epsilon$
photons being scattered by the Ca~II H line out to the Ca~II line at 8662~\AA,
causing this Ca~II absorption line to be {\it filled in} with respect to the
other two Ca~II IR triplet lines.   

We use H$\alpha$ as a proxy for the H$\epsilon$ line.  Following the same
analysis presented by CL97, we determined a relative {\it line  strength}, $F$,
for H$\alpha$ emission and the Ca~II IR triplet absorption lines.  Two points
were selected on either side of the emission or absorption feature.  The
wavelengths  of these points were kept constant for all measurements.  The
observed {\it profile} was integrated across the wavelength window defined by
these two points resulting in an  integrated flux $f_{\ell}$.  A straight line
connected between these two points represent a {\it pseudocontinuum} and the
integrated flux, $f_{c}$, calculated for it.  Then  
 
\begin{displaymath}   
F = \frac{f_{\ell} - f_{c}}{f_{c}},   
\end{displaymath} 
 
\noindent where $F$ will be negative for absorption lines and positive for
emission lines.  The measurements were done for the stars in our sample where
their spectra included wavelengths below 6563~\AA\ and above 8662~\AA; a total
of 27 spectra.  Figure 3 shows the relative line strengths of H$\alpha$, and
the Ca~II $\lambda$8498, Ca~II $\lambda$8542, and Ca~II $\lambda$8662
absorption lines as a function of phase.  The uncertainty in the measurements
is about $\pm$0.007.  The relative line strength of H$\alpha$ clearly shows
large scatter near visible maximum, and is zero within the uncertainty from
phase 0.5 to 0.8.  This is consistent with Balmer emission lines becoming
prominent near maximum visible light.  The Ca~II $\lambda$8498 line strengths
are zero, within the uncertainty of the measurements; variation is not observed
in this line.  The Ca~II $\lambda$8542 line does show some scatter near phase
0.0, and is zero after phase 0.2.  The Ca~II $\lambda$8662 line shows more
scatter than either of the other two Ca~II IR triplet lines, particularly near
phase 0.0.
 
From Figure 3, it is difficult to see any correlation between 
the relative line 
strength of the H$\alpha$ and Ca~II IR triplet lines.  However, we can plot the
relative line strengths of Ca~II IR triplet lines versus the strength of the
H$\alpha$ line to look for correlations.  Figure 4 shows the relative line
strengths of Ca~II $\lambda$8498, Ca~II $\lambda$8542, and Ca~II $\lambda$8662
versus the relative line strength of H$\alpha$.  The data of each plot is
linearly fit and the results of the fits are drawn in the plots.  The slopes
of the Ca~II $\lambda$8498 and $\lambda$8542 relative line strengths versus
H$\alpha$ are 0.08 and 0.05, respectively.  The slope of the linear fit of the
Ca~II $\lambda$8662 relative line strength versus H$\alpha$ relative line 
strength, on the other hand, is 0.32, which is significantly different than the
other two Ca~II IR triplet linear fits.  Furthermore, the Ca~II $\lambda$8662 
versus H$\alpha$ relative line strength slope is positive, so that as the
strength of the H$\alpha$ line increases, the strength of the Ca~II
$\lambda$8662 line decreases (i.e. becomes more positive) and even goes into
emission.  This is the effect we expect if H$\epsilon$ photons are being
scattered by the Ca~II H line out to the Ca~II line at 8662~\AA, causing the
Ca~II $\lambda$8662 absorption line to be filled in with respect to the other
two Ca~II IR triplet lines.

\section{Conclusion}  

The 6200~\AA\ to 9000~\AA\ spectra of Mira variables taken at different phases
support a possible anticorrelation between H$\alpha$ emission and Ca~II
$\lambda$8662 absorption as first suggested by CL97.  Assuming that the
H$\epsilon$ line strength variations are in phase with H$\alpha$, then the
apparent anticorrelation between the strength of the H$\alpha$ emission line 
and the strength of the Ca~II $\lambda$8662 line suggests that a {\it
fluorescence} is taking place in the Ca~II $\lambda$8662 line with H$\epsilon$
serving as the pump through the Ca~II H line.  This type of fluorescence is
common in Mira type variables.  The strong Fe~I (42) lines at 4202~\AA\ and
4308~\AA\ seen in  Miras are well known fluoresced features; in this case, the
ultraviolet Mg~II h \& k lines serve as the pump via an Fe~I (UV3) transition
(e.g., Bidelman \& Herbig 1958; Willson 1976; Luttermoser 1996).
 
The next phase of this research program is to systematically determine 
$T_{\mbox{eff}}$ and $\log g$ as a function of phase for the Miras in our   
sample.  A detailed description of the LTE model synthetic spectra is given by 
Piontek \& Luttermoser (1999). 

ACKNOWLEDGEMENTS.  MWC greatly appreciates support from NSF Grant AST-9500756 
which was the primary source of support this research.  The long-term
observations required for this research project would not be possible without
the continued commitment of the Southeastern Association for Research in
Astronomy to provide the periodic observing times on a meter-class telescope at
Kitt Peak.  We thank Dr.\ Peter Mack of Astronomical Consultants and Equipment
for his excellent technical expertise at the SARA Observatory.  We also thank
Robert Miller at Appalachian State University who machined the spectrograph
adapter so we could use the spectrograph on the DSO 0.45-m telescope. We are
indebted to Marie Rinkoski, NSF sponsored SARA REU student during the Summer
2000, who kindly calculated the phases while reducing a new set of 
spectra as part of her research.  We also greatly appreciate the comments
of the referee of this paper, resulting in several significant improvements.
 In this research, we have used, and 
acknowledge with thanks, data from the AAVSO International Database, 
based on observations submitted to the AAVSO by variable star
observers worldwide.
This research has made use of the SIMBAD databases, operated at 
CDS, Strasbourg, France. 


\section{References}         

\begin{description}        

\item{Bessell, M.S., Scholz, M. \& Wood, P.R. 1996, A\& A, 307, 481} 

\item{Bidelman, W.P. \& Herbig, G.H. 1958, PASP, 70, 451}   

\item{Bowen, G.H. 1988, ApJ, 329, 299}

\item{Brown, J.A., Johnson, H.R., Alexander, D.R., Cutright, L., Sharp, C.M.
1989, ApJS, 71, 623}

\item{Castelaz, M. W. \& Luttermoser, D. G. 1997, AJ, 114, 1584} 

\item{Celis S., L. 1984, AJ, 89, 527}   

\item{Crowe, K., Heaton, B., Castelaz, M. W. 1996, IAPPP Communications, 68,
30}

\item{Contadakis, M. E., \& Solf, J. 1981, A\& A, 101, 241}

\item{Gillet, D. 1988, A\&A, 192, 206}   

\item{Gillet, D., Maurice, E., \& Baade, D. 1983, A\& A, 128, 384}

\item{Gillet, D., Maurice, E., Bouchet, P., \& Ferlet, R. 1985, A\& A, 148, 148}

\item{Haniff, C. A., Ghez, A. M., Gorham, P. W., Kulkarni, S. R., Matthews, K.,
\& Neugebauer, G. 1992, AJ, 103, 1662}

\item{H\"{o}fner, S., J\o rgenson, U.G., Loidl, R., \& Aringer, B. 
1998, A\&A, 340, 497}

\item{Joy, A. H. 1926, ApJ, 63, 281}

\item{Joy, A. H. 1947, ApJ, 106, 288}

\item{Joy, A. H. 1954, ApJS, 1, 39}

\item{Kurucz, R. L. 1970, Smithsonian Ap.\ Obs.\ Special Rept., No.\ 309}

\item{Labeyrie, A., Koechlin, L., Bonneau, D., Blazit, A., \& Foy, R. 1977, ApJ
Letters, 218, L75}   

\item{Lockwood, G. W. 1973, ApJ, 180, 845}   

\item{Lockwood, G. W. \& Wing, R. F. 1971, ApJ, 169, 63}

\item{Loidl, R., H\"{o}fner, S., J\o rgenson, U.G., \& Aringer, B. 
1999, A\&A, 342, 531}

\item{Luttermoser, D.G. 1996, in Ninth Cambridge Workshop on Cool Stars,   
Stellar Systems, and the Sun, R. Pallavicini \& A.K. Dupree, ASP Conference   
Series, 109, 535}   

\item{Luttermoser, D. G. \& Bowen G. H. 1992, in Seventh Cambridge Workshop   
on Cool Stars, Stellar Systems, and the Sun, M.S. Giampapa \& J.A. Bookbinder, 
ASP Conference Series, 26, 558}   

\item{Mattei, J. A., 1996, AAVSO Bulletin 59, Predicted Dates of Maxima        
and Minima of Long Period Variables for 1996}   

\item{Merrill, P. W. 1940, in Spectra of Long-Period Variable Stars,        
(Univ. of Chicago Press; Chicago), p. 44}   

\item Pickering, E. C. 1887, Nature, 36, 32

\item{Piontek, R. \& Luttermoser, D. L. 2000, in Bull.\ of the AAS, 31, 1238} 

\item{Turnshek, D. E., Turnshek, D. A., Graine, E. R. \& Boeshaar, P.C.
1985, An Atlas of Digital Spectra of Cool Stars, (Western Research
Corp.; Tucson, AZ USA)}

\item{Willson, L. A. 1976, ApJ, 205, 172}

\item{Willson, L.A. 2000, Annu.\ Rev.\ Astron.\ Astrophys., 38, 573}   

\item{Wing, R. F. 1967, Ph.D. Thesis, University of California, Berkley}   

\item{Zhou, X. 1991, A\&A, 248, 367}   
\end{description}         

\newpage           

\smallskip         
 
\figcaption{Spectra of the 20 Mira variables.  The name of the star, date, and
phase, are given on each spectrum.  The flux is normalized to one. Above each
set of spectra are markers for the major spectral features, and terrestrial
oxygen.  H$\alpha$ is weak, or not seen in most of the spectra.  Markers
enclose regions of individual spectra where no data was taken.  Spikes
due to cosmic ray hits have not been removed.}

\figcaption{The integrated flux ratio of the 7060-7110~\AA\ wavelength band (A)
to that of the 6995-7045~\AA\ band (B) plotted as a function of phase.  Band A
contains TiO opacity whereas band B is free of TiO, VO, and ZrO bandheads.  
This plot contains only data from the M6, M7, and M8 spectral-type Miras.  No
apparent trends are seen in the data, which indicates that variations in the
H$\alpha$ flux result primarily from intrinsic flux variations in H$\alpha$ and
not from varying overlying TiO absorption.}

\figcaption{Relative line strengths of H$\alpha$ (filled square), Ca~II lines  
at 8498~\AA\ (circle), 8542~\AA\ (triangle), and 8662~\AA\ (diamond), as a  
function of light-variation phase.  Variation in the Ca~II lines at 8498~\AA\ 
and 8542~\AA\ mimic each other, whereas the Ca~II line at 8662~\AA\ does not 
follow the same trend. The uncertainty of the measurements is $\pm$0.007.}   

\figcaption{The relative line strengths of Ca~II $\lambda$8498, Ca~II
$\lambda$8542, and Ca~II $\lambda$8662 versus the relative line strength of
H$\alpha$.  The data of each plot is linearly fit and the results of the fits
are drawn in the plots.  Only the Ca~II $\lambda$8662 versus H$\alpha$ plot
shows a significant slope which implies a correlation between the occurrence of
the two features.}


\include{mwcfigs}
\include{mwcfigsa}
\include{mwcfig34}

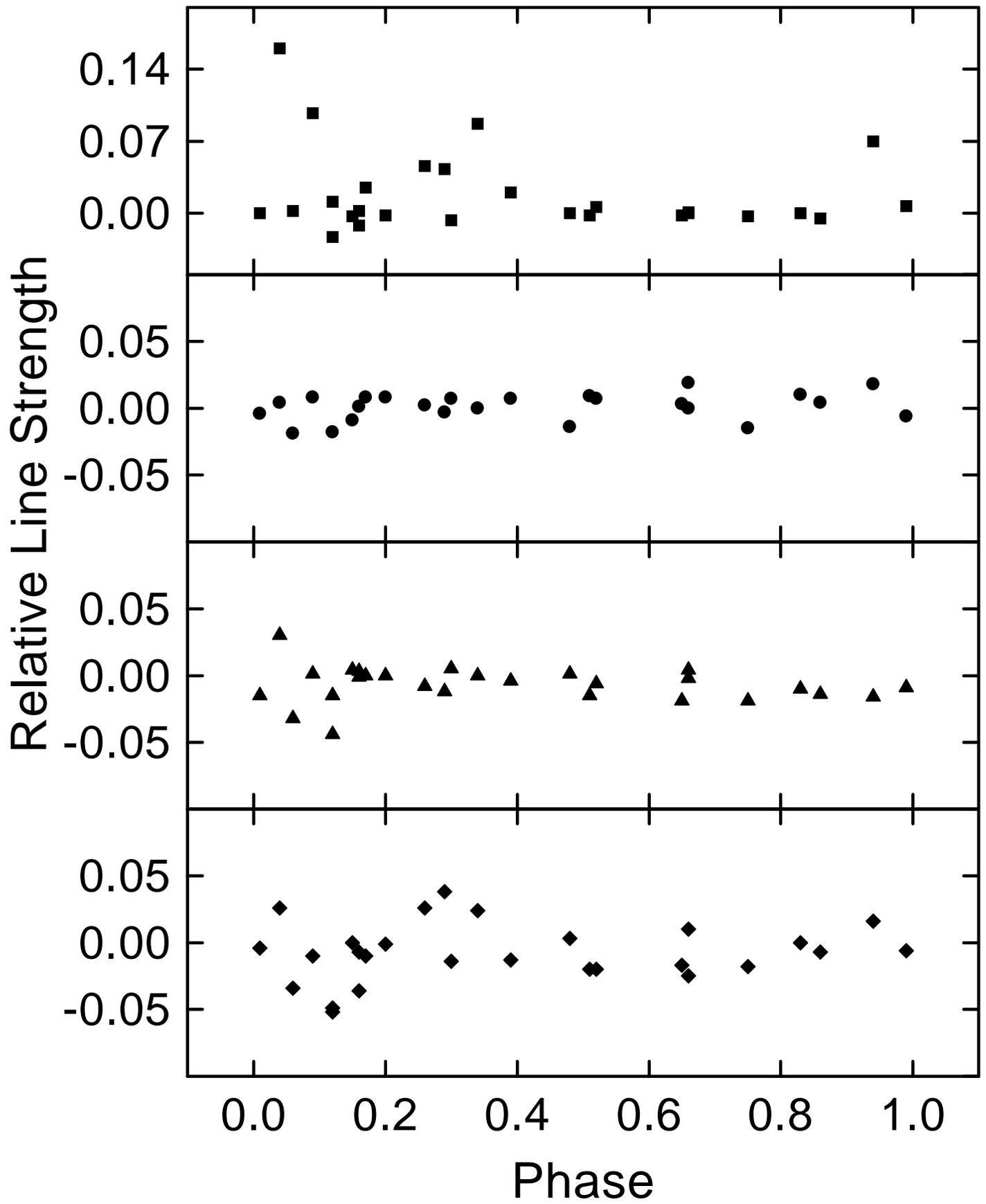
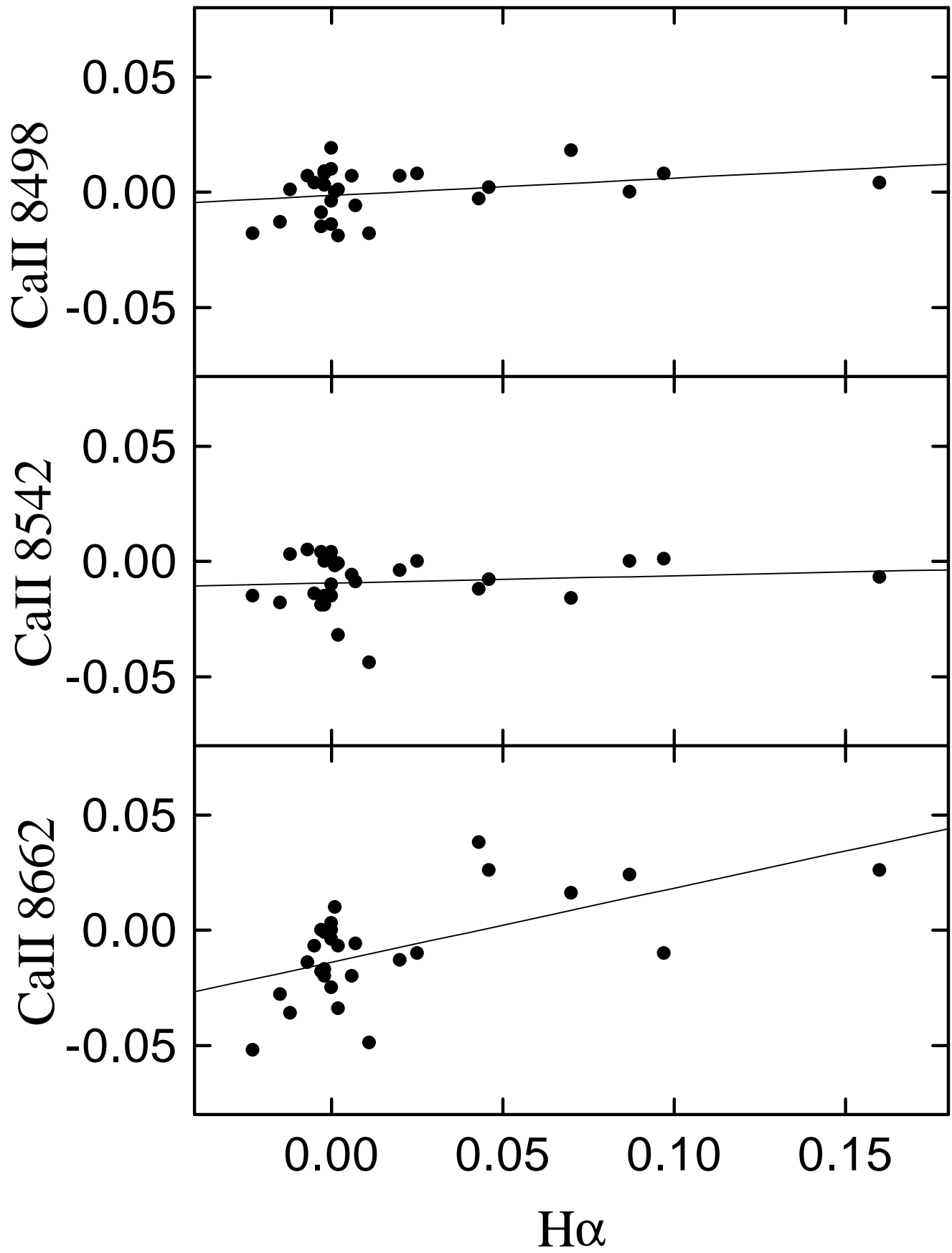
\end{document}


\begin{deluxetable}{lrrcrrl}
\tablenum{2}
\tablewidth{0pc}
\tablecaption{Log of Observations}
\tablehead{\colhead{} & \colhead{Observation} & \colhead{Observation} & 
 \colhead{Visible} & \colhead{} & \colhead{Integration} & \colhead{} \\
\colhead{Star} & \colhead{Date (UT)} & \colhead{JD 2450000+} & \colhead{Phase} 
 & \colhead{V} & \colhead{time (sec)} & \colhead{Telescope}}

\startdata
T Cas   &  1 Nov 97 &  753.5 & 0.04 &  8.0 &  120 & SARA \nl
        & 15 Nov 98 & 1132.5 & 0.92 & 10.2 &  300 & SARA \nl
W And   &  2 Nov 97 &  754.5 & 0.30 & 13.4 &  900 & SARA \nl
$o$ Cet &  1 Nov 97 &  753.5 & 0.80 &  7.1 &   60 & SARA \nl
R Tri   & 15 Feb 97 &  494.5 & 0.73 & 10.0 & 1800 & SARA \nl
        & 13 Mar 97 &  520.5 & 0.83 &  8.6 &  600 & DSO  \nl
        &  1 Nov 97 &  753.5 & 0.70 & 10.3 &  180 & SARA \nl
R Tau   & 15 Feb 97 &  494.5 & 0.16 & 11.7 &  360 & SARA \nl
        &  1 Nov 97 &  753.5 & 0.99 &  8.9 &  180 & SARA \nl
        & 14 Nov 98 & 1131.5 & 0.20 & 11.2 &  300 & SARA \nl
U Ori   & 14 Feb 97 &  493.5 & 0.30 & 10.7 &  600 & SARA \nl
        &  1 Nov 97 &  753.5 & 0.01 &  6.6 &   60 & SARA \nl
        & 15 Feb 98 &  859.5 & 0.29 & 10.5 &  600 & SARA \nl
        & 14 Nov 98 & 1131.5 & 0.03 &  6.4 &  300 & SARA \nl
R Leo   & 14 Feb 97 &  493.5 & 0.16 &  8.5 &  600 & SARA \nl
        & 13 Mar 97 &  520.5 & 0.25 &  9.2 &  600 & DS0  \nl
        & 28 May 97 &  596.5 & 0.49 & 10.0 &  600 & SARA \nl
        & 15 Feb 98 &  859.5 & 0.34 &  9.8 &  300 & SARA \nl
        & 28 Mar 98 &  900.5 & 0.48 & 10.0 &  300 & DSO  \nl
        & 24 May 98 &  957.5 & 0.66 &  9.2 &   60 & SARA \nl
        & 14 Nov 98 & 1131.5 & 0.23 &  8.8 &  300 & SARA \nl
V CVn   & 14 Feb 97 &  493.5 & 0.12 &  7.0 &  240 & SARA \nl
        & 13 Mar 97 &  520.5 & 0.27 &  7.9 &  600 & DSO  \nl
        & 25 May 97 &  593.5 & 0.65 &  8.8 &  600 & SARA \nl
        & 15 Feb 98 &  859.5 & 0.06 &  7.1 &  600 & SARA \nl
        & 23 May 98 &  956.5 & 0.57 &  8.8 &  300 & SARA \nl
R CVn   & 15 Feb 97 &  494.5 & 0.09 &  8.5 &  600 & SARA \nl
        & 13 Mar 97 &  520.5 & 0.17 &  9.5 &  600 & DSO  \nl
        & 26 May 97 &  583.5 & 0.39 & 11.3 & 1800 & SARA \nl
        & 15 Feb 98 &  859.5 & 0.20 &  9.9 &  600 & SARA \nl
        & 23 May 98 &  956.5 & 0.50 & 11.9 &  300 & SARA \nl
V Boo   & 15 Feb 97 &  494.5 & 0.12 &  8.3 &  600 & SARA \nl
        & 26 May 97 &  595.5 & 0.51 & 11.3 &  600 & SARA \nl
        & 24 May 98 &  957.5 & 0.53 &  8.5 &  240 & SARA \nl
R Boo   & 25 May 97 &  583.5 & 0.11 &  8.2 &  600 & SARA \nl
S CrB   & 25 May 97 &  593.5 & 0.48 & 13.6 & 1800 & SARA \nl
        & 24 May 98 &  957.5 & 0.47 & 13.5 &  600 & SARA \nl
R Ser   & 25 May 97 &  593.5 & 0.26 &  6.5 & 1200 & SARA \nl
        & 23 May 98 &  956.5 & 0.27 &  6.9 &  300 & SARA \nl
W Her   & 26 May 97 &  594.5 & 0.86 &  9.8 & 1200 & SARA \nl
        & 24 May 98 &  957.5 & 0.15 & 11.4 &  900 & SARA \nl
RS Her  & 28 May 97 &  596.5 & 0.75 & 11.0 & 1800 & SARA \nl
        & 23 May 98 &  956.5 & 0.40 & 12.3 &  420 & SARA \nl
T Her   & 26 May 97 &  594.5 & 0.15 & 10.3 & 1200 & SARA \nl
        & 24 May 98 &  957.5 & 0.37 & 12.4 & 1200 & SARA \nl
W Lyr   & 26 May 97 &  594.5 & 0.94 &  8.9 &  900 & SARA \nl
$\chi$ Cyg & 26 May 97 & 594.5 & 0.66 & 11.8 & 300 & SARA \nl
        &  1 Nov 97 &  753.5 & 0.04 &  7.6 &   60 & SARA \nl
        & 19 Jul 98 & 1013.5 & 0.66 & 11.3 &  120 & SARA \nl
T Cep   &  2 Nov 97 &  754.5 & 0.52 & 10.2 &  300 & SARA \nl
W Peg   &  1 Nov 97 &  753.5 & 0.13 &  9.0 &  600 & SARA \nl
\enddata
\end{deluxetable}